\documentclass[10pt,twocolumn,letterpaper]{article}

\usepackage{cvpr}
\usepackage{times}
\usepackage{epsfig}
\usepackage{graphicx}
\usepackage{amsmath}
\usepackage{amssymb}
\usepackage{amsfonts}

\usepackage{paralist}

\usepackage{url}

\usepackage[breaklinks=true,bookmarks=false]{hyperref}

\cvprfinalcopy 

\def\cvprPaperID{****} 

1
\begin{document}

\title{Algorithms in Multi-Agent Systems: A Holistic Perspective from Reinforcement Learning and Game Theory}

\author{
Lu Yunlong\\
luyunlong@pku.edu.cn\\
School of EECS, Peking University
\and
Yan Kai\\
289371298@pku.edu.cn\\
School of EECS, Peking University
}
\maketitle

\begin{abstract}
Deep reinforcement learning has achieved state-of-the-art performance in single-agent games, with a drastic development in its methods and applications. Recent works are exploring its potential in multi-agent scenarios. However, they are faced with lots of challenges and are seeking for help from traditional game-theoretic algorithms, which, in turn, show bright application promise combined with modern algorithms and boosting computing power. In this survey, we first introduce basic concepts and algorithms in single-agent RL and multi-agent systems; then, we summarize the related algorithms from three aspects. Solution concepts from game theory give inspiration to algorithms which try to evaluate the agents or find better solutions in multi-agent systems. Fictitious self-play becomes popular and has a great impact on the algorithm of multi-agent reinforcement learning. Counterfactual regret minimization is an important tool to solve games with incomplete information, and has shown great strength when combined with deep learning.  
\end{abstract}


\section{Introduction}

\textit{No man is an island entire of itself.} We live in a world where people constantly interact with each other, and most situations we confronted with every day involve cooperation and competition with others. Though recent years have witnessed dramatic progress in the field of AI, its application will still be limited until intellectual agents have learnt how to cope with others in a multi-agent system. The first attempt of defining multi-agent systems (MAS) dates back to 2000, when Stone and Veloso defined the area of MAS and stated its open problems \cite{paper/early_multiagent_survey}, and the number of algorithms and applications in this area has been rising ever since. In the last decade, multi-agent learning has achieved great success in games which were once considered impossible for AI to conquer, including games with tremendous number of states like Go \cite{paper/alphagozero}, Chess and Shogi \cite{paper/alphazero}, games with incomplete information like Texas Hold’em \cite{paper/poker_deepstack} and Avalon \cite{paper/FFF}, and multi-player video games with extremely high complexity like Dota 2 \cite{blog/openai_dota2} and StarCraft \cite{blog/alphastar}.

The main reason for the current success in multi-agent games is the combination of techniques from two main areas: deep reinforcement learning and game theory. The former provides powerful algorithms for training agents with a particular goal in an interactive environment, but it cannot be straightforwardly applied to a multi-agent setting \cite{paper/station}; the latter is born to analyze the behavior of multiple agents, but is developed mainly in theory, with algorithms capable of solving problems in a very small scale.

Deep reinforcement learning (DRL) is the combination of reinforcement learning (RL) and deep learning. RL \cite{book/rl_introduction} is an area of machine learning concerned with how agents ought to take actions in an environment in order to maximize some notion of cumulative rewards. Deep learning \cite{paper/deep_learning} is a class of machine learning algorithms that uses neural networks (NN) to extract high-level features from the raw input. Before the prevalence of deep learning, RL needs manually-designed features to represent state information when it comes to complex games; neural network can serve as an adaptive function approximator, allowing RL to scale to problems with high-dimensional state space \cite{paper/DQN} and continuous action space \cite{paper/DDPG} \cite{paper/TRPO} \cite{paper/PPO}. 

By adopting such approximator, DRL is proved to be successful in single-player games with high-dimensional input and large state space like Atari \cite{paper/drl_atari}. However, straightforward application of DRL to multi-agent scenarios only achieved limited success both empirically \cite{paper/drl_play_ESS} and theoretically \cite{paper/nonmarkov}. Modeling other agents as part of the environment makes the environment adversarial and no longer Markov, violating the key assumption in the original theory of RL \cite{book/rl_introduction}.

Therefore, a mature approach has to consider carefully the nature of a multi-agent system. Compared with single-agent game where the only agent aims to maximize its own payoff, in a multi-agent system the environment includes other agents, all of whom aim to maximize their payoffs. There are special cases such as potential games or fully cooperative games where all agents can reach the same global optimal; however, under most circumstances an optimal strategy for a given agent does not make sense any more, since the best strategy depends on the choices of others. In game theory \cite{book/watson}, researchers are focused on certain solution concepts, like some kind of equilibrium, rather than optimality \cite{book/multiagent_system}.

Nash equilibrium \cite{book/multiagent_system} is one of the most fundamental solution concepts in game theory, and is widely used to modify single-agent reinforcement learning to tackle multi-agent problems. However, Nash equilibrium is a static solution concept based solely on fixed point, which is limited in describing the dynamic properties of a multi-agent system like recurrent sets, periodic orbits, and limit cycles \cite{paper/cyclic} \cite{paper/alpharank}. Some researchers try to establish new solution concepts more capable of describing such properties, and use them to evaluate or train agents in a multi-agent system. The attempts include bounded rationality \cite{paper/MAAIRL} and dynamic-system-based concepts such as Markov-Conley Chain (MCC), proposed by $\alpha$-rank \cite{paper/alpharank} and applied in \cite{paper/alpharank_train} and \cite{paper/alphaalpha}.

In addition to focusing on solution concepts, some game-theoretic algorithms have sparked a revolution in computer game playing of some of the most difficult games.

One of them is called is called fictitious self-play \cite{paper/fsp}, which is a sample-based variant of fictitious play \cite{paper/brown_original_fp}, a popular game-theoretic model of learning in games. In this model, players repeatedly play a game, at each iteration choosing a best response to their opponents’ average strategies. Then the average strategy profile of fictitious players converges to a Nash equilibrium in certain classes of games. This algorithm has laid the foundation of learning from self-play experiences and has a great impact on the algorithm of multi-agent reinforcement learning, producing many exciting results including AlphaZero \cite{paper/alphazero} when combined with enough computing power. 

Another one of them is called counterfactual regret minimization \cite{paper/cfr_survey}, which is based on the important game-theoretic algorithm of regret matching introduced by Hart and Mas-Colell in 2000 \cite{paper/regret_matching}. In this algorithm, players reach equilibrium play by tracking regrets for past plays, making future plays proportional to positive regrets. This simple and intuitive technique has become the most powerful tool to deal with games with incomplete information, and has achieved great success in games like Poker \cite{paper/poker_deepstack} when combined with deep learning. In this survey, we will focus on the above two algorithms and how they are applied in reinforcement learning, especially in competitive multi-agent RL.

This survey consists of four parts: Section 2 includes the basic concepts and algorithms in single-agent reinforcement learning, as well as basic concepts of multi-agent system; Section 3 mainly describes the inspiration that solution concepts give to multi-agent RL; Section 4 states how fictitious self-play grows into an important tool for multi-agent RL; Section 5 gives an introduction for counterfactual regret minimization and its application in multi-agent RL.

\section{Background}

\subsection{Single-Agent Reinforcement Learning}

One of the simplest forms of single-agent reinforcement learning problems is the multi-armed bandit problem \cite{paper/MAB}, where players try to earn money (maximizing \emph{reward}) by selecting an arm (an \emph{action}) to play (\emph{interact} with the environment). This is an oversimplified model, where the situation before and after each turn is exactly the same. In most scenarios, you may face different conditions and take multiple actions in a row to complete one turn. This introduce the discrimination between \emph{states} and their \emph{transitions}.

More formally, a single-agent RL problem can be described as a Markov Decision Process (MDP) \cite{book/rl_introduction}, which is a quintet $<S, A, R, T, \gamma>$ where $S$ represents a set of states and $A$ represents a set of actions. The transition function $T : S \times A \xrightarrow{} \Delta(S)$ maps each state-action pair to a probability distribution over states. Thus, for each $s, s' \in S$ and $a \in A$, the function $T$ determines the probability of going from state $s$ to $s'$ after executing action $a$. The reward function $R : S \times A \times S \xrightarrow{} \mathbb{R}$ defines the immediate reward that an agent would receive when transits from state $s$ to $s'$ via action $a$; sometimes the reward can also be a probability distribution on $\mathbb{R}$. $\gamma \in [0, 1]$ defines the discount factor to balance the trade-off between the reward in the next transition and rewards in further steps, and the reward gained $k$ steps later are discounted by a factor of $\gamma^k$ in the accumulation.

The solution of MDP is a policy $\pi : S \xrightarrow{} \Delta(A)$, which maps states to a probability distribution of actions, indicating how an agent chooses actions when faced with each state. The goal of RL is to find the \emph{optimal policy} $\pi^{*}$ to maximize the expected discounted sum of rewards. There are different techniques for solving MDPs assuming all of its components are known. One of the most common techniques is the value iteration algorithm \cite{book/rl_introduction} based on the Bellman equation:
\begin{equation}
    v_{\pi}(s)=\mathbb{E}_{a \sim \pi(s)}\mathbb{E}_{s' \sim T(s, a)}[R(s, a, s')+\gamma v_{\pi}(s')]
\end{equation}
This equation expresses the value (expected payoffs) of a state under policy $\pi$, which can be used to obtain the optimal policy $\pi^{*}=arg max_{\pi}v_{\pi}(s)$ i.e. the one that maximizes the value function.

Value iteration is a model-based RL algorithm since it requires a complete model of the MDP. However, in most cases an environment is either not fully known or too complicated. For this reason, model-free algorithms are more preferred, which learn from experiences of interacting with the environment in a sample-based fashion.

There are two kinds of model-free RL algorithms, value-based and policy-based. Value-based algorithms optimize the policy by tracking the values of states or state-action pairs, and choosing better actions according to these values. Q-learning is one of the most well-known of them. A Q-learning agent learns a Q-function, which is the estimate of the expected payoff starting in state $s$ and taking action $a$ as $Q(s, a)$. Whenever the agent transits from state $s$ to $s'$ by taking action $a$ with reward $r$, the $Q$ table is updated as:
\begin{equation}
    Q(s, a) = Q(s, a) + \alpha[(r + \gamma max_{b}Q(s', b)) - Q(s, a)]
\end{equation}
where $\alpha$ is the learning rate. It is proved that Q-learning can converge to the optimal $Q^{*}$ if each state-action pair is visited infinitely often under specific parameters \cite{paper/rl_value_base} \cite{paper/q_learning}. The most famous value-based method in DRL is the Deep Q-network (DQN) \cite{paper/DQN}, which introduces deep learning into RL by approximating the $Q$ function with a deep NN, allowing RL to deal with problems with high-dimensional state space like pixel space. Moreover, since the neural networks are capable of extracting features by themselves, manual feature engineering with prior knowledge is no longer necessary. 

Policy-based algorithms take another route, which directly learn parameterized policies based on gradients of some performance measures using gradient descent method. One of the earliest work is REINFORCE \cite{paper/policy_gradient}, which samples full episode trajectories with Monte Carlo methods to estimate return. The policy parameters $\theta$, where $\pi(a|s, \theta) \approx \pi(a|s)$, is updated as:
\begin{equation}
    \theta_{n+1} = \theta_{n} + \alpha G_{n} \frac{\nabla \pi(A_{n}|S_{n}, \theta_{n})}{\pi(A_{n}|S_{n}, \theta_{n})}
\end{equation}
where $\alpha$ is the learning rate and $G$ is the discounted sum of rewards. However, pure policy-based methods can have high variance \cite{book/rl_introduction} and actor-critic algorithms \cite{paper/actor_critic}, a combination of value-based and policy-based methods, have been proposed. They use actors to learn parameterized policies and critics to learn value functions, which allows the policy updates to take into consideration the value estimates to reduce the variance compared to vanilla policy-based methods.

Combining policy-based methods with deep learning is straightforward since the parameterized policy can be directly replaced with a deep NN, and there are many variants of them. Deep deterministic policy gradient (DDPG) \cite{paper/DDPG} addresses the issue of large action space by adding sampled noise, such as noise drawn from Ornstein-Uhlenbeck process \cite{paper/ornstein}, to its actor's policy, allowing more exploratory behavior. Asynchronous Advantage Actor-Critic (A3C) \cite{paper/A3C} is a distributed algorithm where multiple actors running on different threads interact with the environment simultaneously and compute gradients in a local manner. UNREAL framework \cite{paper/UNREAL} is based on A3C and proposes unsupervised auxiliary tasks like reward prediction to accelerate the learning process. Importance Weighted Actor-Learner Architecture (IMPALA) \cite{paper/IMPALA} is another distributed algorithm that allows trajectories of experience to be communicated between actors and a centralized learner. Trust Region Policy Optimization (TRPO) \cite{paper/TRPO} and Proximal Policy Optimization (PPO) \cite{paper/PPO} are state-of-the-art policy-based DRL algorithms, where changes in policy are incorporated to the loss function by adding KL-divergence to the loss to prevent abrupt changes in policies during training.

\subsection{Markov Game and Nash Equilibrium}

While MDP models single-agent game with discrete time, the framework of Markov games, or stochastic games \cite{paper/early_stochastic_games} , models multi-agent systems with discrete time and non-cooperative nature. An n-player Markov game is defined \cite{book/formal_stochastic_games} by a tuple $<S, A^1, ..., A^n, r^1, ..., r^n, p>$, where $s$ is the state space, $A^i$ is the action space of player $i$, $r^i : S \times A^1 \times ...\times A^n \xrightarrow{} R$ is the payoff function for player $i$, $p : S \times A^1 \times ...\times A^n \xrightarrow{} \Delta(S)$ is the transition probability map, where $\Delta(S)$ is the set of probability distributions over state space $S$. Given state $s$, agents independently choose actions $a^1, ..., a^n$, and receive rewards $r^i(s, a^1, ..., a^n)$, $i=1, ..., n$. The state then transits to the next state based on the transition probabilities. Specifically, in a discounted Markov game, the objective of each player is to maximize the discounted sum of rewards, with discount factor $\gamma \in [0, 1)$. Let $\pi^i$ be the strategy of player $i$, for a given initial state $s$, player $i$ tries to maximize
\begin{equation}
    v^i(s, \pi^1, ..., \pi^n)=\sum_{t=0}^{\infty}\gamma^{t}\mathbb{E}(r_{t}^i | \pi^1, ..., \pi^n, s_{0}=s)
\end{equation}

Nash equilibrium is one of the most fundamental solution concepts in game theory. It is a joint strategy profile where each agent's strategy is a best response of the others'. Specifically, for a Markov game, a Nash equilibrium is a tuple of n strategies $(\pi_{*}^1, ..., \pi_{*}^n)$ such that for all state $s \in S$ and $i = 1, ..., n$,
\begin{equation}
\begin{aligned}
    v^i(s, \pi_{*}^1, ..., \pi_{*}^n) \ge
    v^i(s, \pi_{*}^1, ...,\pi_{*}^{i-1} ,\pi^i ,\pi_{*}^{i+1} , ..., \pi_{*}^n)
    \\for \; all \; \pi^i \in \Pi^i
\end{aligned}
\end{equation}
where $\Pi^i$ is the set of all strategies available to agent i.

In general, the strategies that constitute a Nash Equilibrium can be either stationary strategies or behavior strategies which allow conditioning of actions on history of play. In 1964 Fink \cite{paper/nash_eq_markov_game} proved that every $n$-player discounted stochastic game possesses at least one Nash equilibrium point in stationary strategies, as a enhanced conclusion of Nash's theorem in \cite{paper/nash_eq}.

\section{Solution Concepts}

\subsection{Without Solution Concepts: Independent RL}

Although algorithms for single-agent reinforcement learning are not designed for multi-agent systems because the assumptions from which they are derived are invalid, there are a group of works studying and analyzing the behaviors directly using independent DRL agents in multi-agent settings, modeling other agents as part of the environment.

Tampuu \etal \cite{paper/rl_multiagent_DQN_pong} trained two independent DQN learning agents to play the Pong game. They tried to adapt the reward functions to achieve either cooperative or competitive settings. Later, Leibo \etal \cite{paper/rl_multiagent_DQN_social} also tried independent DQNs but in the context of sequential social dilemmas. This work showed that cooperative or competitive settings can not only affect discrete actions, but also change the whole policies of agents.

Recently, Bansal \etal \cite{paper/rl_multiagent_PPO_mojoco} trained independent learning agents with PPO using the MuJoCo simulator \cite{paper/mojoco}. They applied two modifications to deal with the multi-agent nature of the problem though. First, they used exploration rewards which are dense rewards to allow agents to learn basic, non-competitive behaviors, and reducing this type of reward through time, giving more weight to the environmental, competitive reward. Second, they maintained a collection of older versions of the opponent to sample from, rather than always using the most recent version, to stabilize the behavior changes over the training process.

Raghu \etal \cite{paper/drl_play_ESS} investigated how different DRL algorithms, including DQN, A2C and PPO, performed in a family of two-player zero-sum games with tunable complexity, called Erdos-Selfridge-Spencer games, which is a parameterized family of environments and the optimal behavior can be completely characterized. Their work showed that different DRL algorithms can show wide variation in performance as the game's difficulty is tuned.

\subsection{RL towards Nash Equilibrium}

To tackle multi-agent problems, directly using single-agent reinforcement learning proves invalid, and efforts are made to better understand the nature of multi-agent systems. The first step is taken by Littman in 1994 \cite{paper/Minimax-Q}, who pointed out that for many Markov games, there is no policy that is undominated because performance depends critically on the choice of opponent. Instead, each policy should be evaluated with respect to the opponent that makes it look the worst. He proposed Minimax-Q in the setting of two-player zero-sum Markov game, where the opponent is not modeled as part of the environment but one that rationally makes your policy worse. Minimax-Q is a modification of single-agent Q-learning, where the original $Q(s, a)$ is substituted with $Q(s, a, o)$, representing the expected payoff for taking action $a$ when the opponent chooses action $o$ from state $s$ and continuing optimally thereafter. The updating rule becomes:
\begin{equation}
\begin{aligned}
    Q(s, a, o) = Q(s, a, o) + \alpha[(r + \gamma V(s'))) - Q(s, a, o)]
    \\where \; V(s) = \max_{\pi \in \Delta(A)}\min_{o \in O}\mathbb{E}_{a \sim \pi}Q(s, a, o)
\end{aligned}
\end{equation}

Then in 2001, he proposed Team-Q \cite{paper/Team-Q} to tackle another special case of Markov games called team Markov game, where agents share the same reward function i.e. have the same objective. In this case, $Q_i(s, a^1, ..., a^n)$ is the Q-function of agent $i$ and the updating rule becomes:
\begin{equation}
\begin{aligned}
    Q_i(s, \vec{a}) = Q_i(s, \vec{a}) + \alpha[(r + \gamma V_i(s'))) - Q_i(s, \vec{a})]
    \\where \; V_i(s) = \max_{\vec{a} \in A(s)}Q_i(s, \vec{a})
\end{aligned}
\end{equation}
In this work, he pointed out that Minimax-Q and Team-Q can be considered as two special cases of a more general algorithm called Nash-Q, because both algorithms are updating the Q-function with values from Nash equilibrium, either in cooperative or competitive scenarios. In the same year, he unified them into Friend-and-Foe-Q \cite{paper/FF-Q}, where cooperation equilibrium and adversarial equilibrium can co-exist in a multi-agent system, and the update of Q-function is unified into:
\begin{equation}
\begin{aligned}
    Q_i(s, \vec{a}) = Q_i(s, \vec{a}) + \alpha[(r + \gamma V_i(s'))) - Q_i(s, \vec{a})]
    \\\text{where} \; V_i(s) = \max_{\pi \in \Delta(X_1) \times ... \times \Delta(X_{k})}\min_{y_1, ..., y_{l} \in Y_1 \times ... \times Y_{l}}
    \\\mathbb{E}_{x_1, ..., x_{k} \sim \pi}Q_i(s, x_1, ..., x_{k}, y_1, ..., y_{l})
\end{aligned}
\end{equation}
where $X_1, ..., X_{k}$ are the actions available to the $k$ friends of player $i$ and $Y_1, ..., Y_{l}$ are those of $l$ foes, based on the idea that $i$'s friends are working together to maximize $i$'s value, while $i$'s foes are working together to minimize $i$'s value.

Although Littman had suggested a general Nash-Q algorithm in \cite{paper/Team-Q}, it was not until Hu \etal \cite{paper/Nash-Q} that a formal formulation of Nash-Q and poof of convergence were proposed. In this work, the Q-function is updated by:
\begin{equation}
\begin{aligned}
    Q_i(s, \vec{a}) = Q_i(s, \vec{a}) + \alpha[(r + \gamma Nash_i(s'))) - Q_i(s, \vec{a})]
\end{aligned}
\end{equation}
where $Nash_i(s)$ is agent $i$'s payoff in state $s$ in the selected Nash equilibrium. It is proved that Nash-Q converges under a strict condition that, every stage game during learning has either a global optimal or a saddle point, and they are always selected to update the Q-function, where a saddle point means a Nash equilibrium where each agent would receive a higher payoff when at least one of the other agents deviates.

Shortly after, multiple variants of Nash-Q were proposed in a very similar fashion. Greenwald \cite{paper/Correlated-Q} used a general form of Nash equilibrium called correlated equilibrium \cite{paper/correlated-EQ}, which is a probability distribution over the joint space
of actions, where all agents optimize with respect to others' probabilities, conditioned on their own. He proposed Correlated-Q and its four variants focused on different correlated equilibrium, as an attempt to unify the previous works:
\begin{enumerate}
    \item maximize the sum of the players’ rewards
    \item maximize the minimum of the players’ rewards
    \item maximize the maximum of the players’ rewards
    \item maximize the maximum of each individual player $i$’s rewards
\end{enumerate}
In the same year, K\"{o}n\"{o}nen \cite{paper/Asymmetric-Q} introduced Asymmetric-Q to deal with Stackelberg leadership model, a game where the leader moves first and the follower move sequentially. Based on the hierarchical equilibrium solution concept called Stackelberg equilibrium \cite{paper/Stackel}, Asymmetric-Q uses a updating rule of Q-function similar to Nash-Q, except in a hierarchical learning order.

\subsection{Beyond Nash Equilibrium}

Nash Equilibrium proves to be powerful in several specific types of games, but it has never been compatible with general-sum games with more than two players. When Nash equilibrium is applied to reinforcement learning, small changes in the values of joint-actions may cause a large change in the state’s Nash equilibria, making the training unstable and unable to converge \cite{paper/00ICML_MIKE}. There are attempts, both theoretically and experimentally, to bypass the incompatibility by either redefining the equilibrium \cite{paper/LNEGSMG} or straightforwardly applying the algorithm for two-player zero-sum games to general-sum games with more than two players \cite{paper/manyplayerpoker}, but these attempts only provide limited success in specific scenarios.

Another problem is that Nash equilibrium is notoriously hard to compute. Chen and Deng \cite{paper/2pNPPADC} proved that calculating Nash equilibrium for general-sum games is PPAD-complete for two players, which requires exponential-time algorithm. Even if all the equilibria are found, there is no guarantee which equilibrium an algorithm would fall in; it may probably find an ineffective equilibrium with less payoff for both agents. There are researches about other kinds of equilibrium such as Stackelberg equilibrium \cite{paper/Stackel} \cite{paper/Asymmetric-Q}. However, these equilibrium can only deal with specific types of problems, far less useful than Nash equilibrium defined for general scenarios. Therefore, current studies about game theory in RL are still limited in the few types of games Nash can solve, including two-player zero-sum games and team Markov games.

Recently, researchers from DeepMind proposed an algorithm called $\alpha$-rank \cite{paper/alpharank} for ranking the performance of different agents in a multi-agent system. In this work, they introduced a new solution concept called Markov-Conley Chain (MCC), based on strongly sink connected components on transition graphs of Markov chains, which can be seen as a discrete version of dynamic systems in continuous space. According to the Conley's theorem \cite{book/Conley}, the domain of any dynamic system can be decomposed into its chain components with the remaining points transient, leading to the recurrent part by a complete Lyapunov function. Therefore, under the solution concept based on dynamic systems, there exists a potential function, which is a complete Lyapunov function, as the incentive of optimization for all agents. Such games, called potential games, have Nash equilibrium at local optima, thus greatly simplifying the process of finding equilibrium.

$\alpha$-rank is inspired by evolutionary algorithms. Consider a monomorphic population wherein all individuals play identical strategies, and a monomorphic population profile consisting multiple populations where each population may be playing a different strategy. Mutation rate is fixed to a very small number so that each time there is at most one individual in one population that is mutated. Individuals are constantly chosen from each population to involve in the game and be evaluated and those with better performance have better chance to reproduce. So the only mutation will either spread until taking over the whole population, or be wiped out.

Since the result of mutation only ends up in changing behavior of the whole population or not, this process can be viewed as a random walk of the whole monomorphic population profile on states of different strategy profiles, where adjacent states differ by exactly one agent's strategy. This graph is called "response graph". More specifically, the transitional matrix of this Markov process can be written as:
\begin{equation}
\rho_{s_i^k,s_j^k}^{k}(s^{-k})=\left\{\begin{aligned}
&\frac{1-e^{-\alpha(r^k_i-r^k_j)}}{1-e^{-m\alpha(r^k_i-r^k_j)}}\ \text{if }r^k_i\neq r^k_j\\
&\frac1{m}\qquad\qquad\qquad\ \text{if }r^k_i=r^k_j\\
\end{aligned}\right.
\end{equation}
\begin{equation}
C_{i,j}=\left\{\begin{aligned}
&\eta\rho_{ s_i^k,s_j^k}^k(s^{-k})\ \text{if }\exists k \text{ such that } s_i^k\neq s_j^k\text{ and }s_i^{-k}=s_j^{-k}\\
& 1-\sum_{j\neq i}C_{i,j} \ \ \text{if}\ s_i=s_j\\
& 0\qquad\qquad\quad\text{otherwise}\\
\end{aligned}\right.
\end{equation}
where $C_{i,j}$ is the transition probability from strategy profile $i$ to profile $j$, $r_i^k$ and $r_j^k$ are the rewards of agent $k$ in profile $i$ and $j$, $s_i^{k}$ is $k$-th player's strategy in profile $i$ and $s_i^{-k}$ is others' strategies in profile $i$, $\eta$ is the reciprocal of adjacent strategy profile, $\alpha$ is a hyper-parameter standing for the harshness of natural selection and $m$ is the number of individuals in each population.

After constructing such Markov chain, its stationary distribution can be calculated and the larger probability a profile has, the stronger and more robust it is, which can be used to rank different strategy profiles. When $\alpha$ goes to infinity, all probability on the graph will eventually converge to the sink strong connected component of the directed response graph, where one profile is connected to another if and only if there exists a player with utility improvement. Therefore, the complete Lyapunov function guaranteed by Conley's theorem serves as an approximate potential function on the response graph, which is the theoretical basis of the algorithm.

The invention of a new solution concept and its corresponding ranking algorithm provides a powerful polynomial-time tool for strategy analysis of multi-agent general-sum games, bypassing the limitations of Nash equilibrium and surpassing Elo, which can perform badly when there is no total-order relations or there is a large population of low-level agents. Later, Torreno \etal \cite{paper/MEII} generalizes this algorithm to partially observable environments. However, researchers from Huawei \cite{paper/alphaalpha} claim that $\alpha$-rank is impractical, for they use strategy profiles to conceal the exponential essence of their algorithms with respect to the number of strategies and agents. Instead, they propose a scalable alternative called $\alpha^\alpha$-rank based on MCC, which is capable of evaluating tens of millions of strategy profiles. Their key idea is to use sample-based stochastic optimization instead of a linear-algebra-based solution, and solving the stationary distribution is rewritten as an optimization problem to minimize the distance of a vector and the vector multiplied by the distribution matrix in this case.

\section{Fictitious Self-Play}

\subsection{Classical Fictitious Play}
Fictitious play is first proposed by Brown in 1951 \cite{paper/brown_original_fp}. In fictitious play, each agent models other agents' average strategies, and play the best response against them. The process is repeated until convergence if possible. It is proved that the original fictitious play can converge to Nash equilibrium for two-player zero-sum games \cite{paper/Robinson1951}, 2*2 games \cite{paper/Miyasawa1961}, potential games \cite{paper/Shapley1996}, games with an interior ESS \cite{paper/Hofbauer1995} and certain classes of super-modular games \cite{paper/Hahn1999}.

\begin{figure}[t]
\begin{center}
  \includegraphics[width=0.8\linewidth]{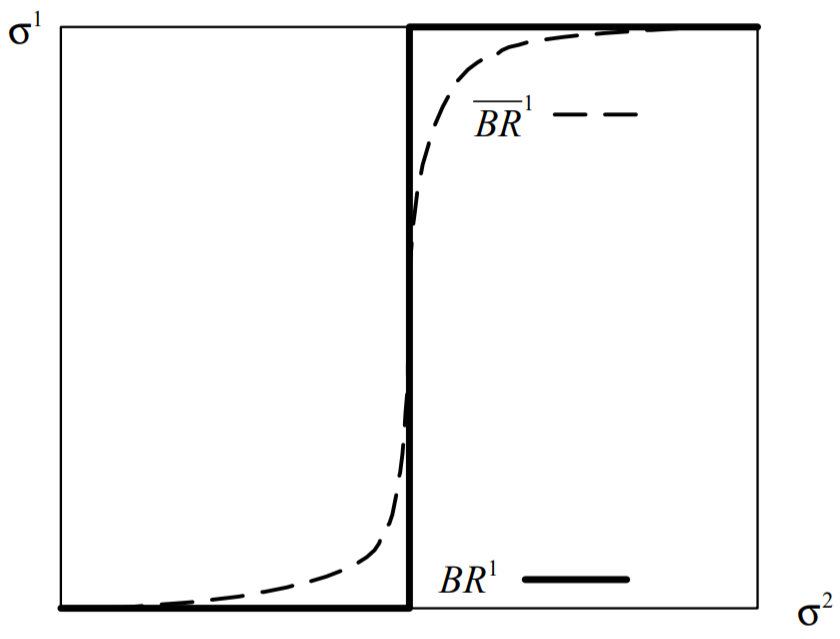}
\end{center}
   \caption{The best response has been “smoothed" into a continuous function by introducing noise. BR stands for best response and $\delta_1, \delta_2$ are for two strategies. The picture is from \cite{book/TTLG}.}
\label{fig:best_response}
\end{figure}

However, since best response is generically pure, a player's choices under fictitious play are sensitive to the values of the opponents' average strategy, and experiments can never converge to a mixed Nash equilibrium (even if players have a mixed best response in mind). Worse still, fictitious play does not provide convergence guarantee for general sum games. Therefore, Fudenberg and Kreps introduced stochastic fictitious play \cite{paper/Kreps1993}. Their setting assumes that in a standard game where players move simultaneously, each player privately observes a small noise on their payoffs. This noise turns the original game into a stochastic game and smooths the best response, turning it into a continuous function, as is shown in Fig. \ref{fig:best_response}.

The benefits are threefold. First, by creating a continuous function, we can correspond the (possibly mixed) equilibrium of the original game with each Nash equilibrium of the perturbed game. Second, such uncertainty prevents the player's strategy from being deduced and exploited by others. Third, this is a model more consistent with actual human psychology \cite{book/TTLG}.

In their work, it is proved that under some assumptions any mixed-strategy equilibrium of the original game is a limit of pure-strategy Bayesian equilibrium of the perturbed game, when the noise approaches zero \cite{book/TTLG}. Later, Perkins \etal \cite{paper/sfpwcas} further proved that stochastic fictitious play converges to an equilibrium in two-player zero-sum games and its variant converges to an equilibrium in negative-definite single-population games with continuous action sets.

For games that admit a unique Nash equilibrium, the empirical distribution of stochastic fictitious play almost surely converge to the Nash equilibrium. It is later proved by \cite{paper/OGC} that four special types of games can converge globally: games with an interior ESS (Evolutionary Stable Strategy), zero-sum games, potential games and all super-modular games. Super-modular games are a special type of games where the marginal value of one player's action will increase due to other players' actions.

\subsection{RL with Fictitious Self-Play}
There are two main obstacles preventing fictitious play from incorporating reinforcement learning. For one thing, fictitious play is an algorithm that requires playing best response to others' average strategies in each stage, while reinforcement learning is a learning algorithm that optimizes the policy, slowly moving to the best response. For another, fictitious play is designed on normal-form games, while reinforcement learning is modeled on MDPs, which can be seen as extensive games with multiple steps. Rewriting an extensive game in matrix form is extremely inefficient because it can yield exponentially large state and action spaces.

Genugten \cite{paper/WFFP} unintentionally took the first step of combining fictitious play and reinforcement learning, when he proposed weakened fictitious play in two-player zero-sum games, where agents play a better response to others' average policy instead of best response. In this work, he introduced $\epsilon$-best response, the set of response with reward $r$ so that $r \le R-\epsilon$ when $R$ is the best response, where $\epsilon$ is initially large but asymptotically vanishing to 0. Weakened fictitious play is originally invented for speeding up the process of convergence, but it unintentionally opens the door for machine learning algorithms like reinforcement learning, which slowly optimizes the policy towards best response. Later, Leslie \etal \cite{paper/GWFP} generalized the setting from two-player zero-sum games to general-sum games with more than two players.

Then in 2015, Heinrich \etal \cite{paper/fsp} proved a critical lemma based on Kuhn's theorem \cite{paper/kuhn1953} that for a player with perfect recall, any mixed strategy is realization-equivalent to a behavior strategy, and vice versa. Heinrich's lemma gives the method of mixing normal-form strategies by a weighted combination of realization-equivalent behavior strategies. The formal statement is as follows: Let $\pi$ and $\beta$ be two behavior strategies, $\Pi$ and $B$ be two mixed strategies that are realization-equivalent to $\pi$ and $\beta$, and $\lambda_1, \lambda_2\in R_{\geq 0}$ with $\lambda_1+\lambda_2=1$. Then for each information state $u\in U$,
\begin{equation}
    \mu(u)=\pi(u)+\frac{\lambda_2 x_\beta(\delta_u)}{\lambda_1 x_\pi(\delta_u)+\lambda_2 x_\beta(\delta_u)}(\beta(u)-\pi(u))
\end{equation}
defines a behavioural strategy $\mu$ and $\mu$ is realization equivalent to the mixed strategy $M = \lambda_1\Pi + \lambda_2 B$.

This formula expresses mixed strategies in extensive-form games without rewriting them into normal-form games. By using deep reinforcement learning (DQN in his work) and imitation learning for the calculation of best response and average policies, Heinrich proposed Neural Fictitious Self Play (NFSP) \cite{paper/fsp} in the same paper, marking the beginning of fictitious self-play entering multi-agent reinforcement learning. In 2016, Heinrich generalized NFSP to partially observable environments, achieving superhuman performance on limited Texas Poker \cite{paper/SPII}. It is worth noting that deep reinforcement learning serves not only as a powerful calculation method, but also as an end-to-end solver eliminating the need to manually design features with prior knowledge in specific fields.

More attempts are made to combine RL algorithms with fictitious play. Zhang \etal \cite{paper/MC-NFSP} proposes MC-NFSP, which uses Monte Carlo tree search to further enhance the best strategy. In the tree search, the agents choose action to maximize a weighted sum of Q-function and an exploration reward, thus enhancing the exploration ability of the self-play. The paper also proposes Asynchronous NFSP (ANFSP), which is an extension of original NFSP to allow distributed training. Perolat \etal \cite{paper/ACFP} builds a stochastic approximation of the fictitious play process with an architecture inspired by actor-critic algorithms. In this algorithm, the actor and critic updates their policy together towards best response, and it is proved that in a zero-sum two-player multi-state game, this algorithm converges to a Nash equilibrium.

\subsection{Applications of Fictitious Self-Play}

There are many applications of fictitious self-play in reinforcement learning, mostly related to adversarial training and evolutionary algorithms. Gupta \etal \cite{paper/ARLODASCA} use self-play to simultaneously train a cyber attacker and defender. The attacker attempts to deform the input from ground truth to reduce the performance of an RL agent, while the defender functions as a preprocessor for that agent, correcting the deformed input. LOLA algorithm \cite{paper/LOLA}, proposed by OpenAI, takes the opponents' gradient into consideration in adversarial training, and achieves fancy results using self-play on complicated environments such as MuJoCo \cite{paper/mojoco}. Kawamura \etal \cite{paper/ELF} empirically combine policy gradient algorithms with NFSP and successfully apply it on a non-trivial RTS game. AlphaGo and its successors \cite{paper/alphazero} combine self-play with deep reinforcement learning and Monte Carlo tree search, achieving state-of-the-art performance on two-player zero-sum games like Go, Chess and Shogi. Sukhbaatar \etal \cite{paper/IMAC} use self-play to conduct an unsupervised curriculum learning, letting the agents learn about its environment without giving a reward function.

Double oracle \cite{paper/doubleoracle} is a well-known variant of fictitious self-play on two-player zero-sum games, which is designed for adapting environment with a reward function correlated with the opponent's behavior. In the algorithm, the "oracle" stands for the mechanism to calculate the best pure strategic response of any mixed strategy of both sides. The algorithm starts from a small set of actions, and both sides calculate the optimal mixed strategy against the previous set. Then, both sides assume that the opponent will take the optimal mixed strategy, and calculate the best pure strategic response against them, which is added to the set. The process is repeated until the set no longer grows. Bo\v{s}ansk\'{y} \cite{paper/DOEX} extends double oracle to extensive-form games, where the agents choose a sequence of actions from a set of action sequences, and the best response sequence is added for next iteration. Double oracle is proved to converge to a minimax equilibrium for two-player zero-sum games, and has applications in security scenarios \cite{paper/ZOSG}.

Policy Space Response Oracle (PSRO) \cite{paper/PSRO} is a grand unified model of fictitious play, independent reinforcement learning and double oracle, for all of them are specific instances of PSRO. PSRO is built on meta-game analysis, inherited from empirical game-theoretic analysis (EGTA) \cite{paper/EGTA}. For every specific game to solve, there exists a meta-game whose action is to choose a policy. A finite set of policies is maintained as a population, which grows as more policies are explored later in training. For each agent in each episode, a set of opponents' policies is sampled from the population and a best response (the oracle) is calculated, which will be included in the population in the next iteration. Final output of the algorithm is some kind of combination of the population. As for fictitious play, the final output is the average policy of the entire population; as for independent reinforcement learning, the output is the last policy of the population. One of the most successful application of PSRO is AlphaStar by DeepMind \cite{blog/alphastar}, which achieves professional level in the multi-player video game StarCraft.

As a meta-game solver, PSRO is compatible with high-level solution concept. Recently, there are attempts \cite{paper/alpharank_train} to combine PSRO with $\alpha$-rank \cite{paper/alpharank} to get theoretical proof of performance for general-sum games with multiple agents. In a high-level perspective, PSRO iteratively calculates the best response and puts it in the population, which is exactly the process where probabilities "flow" on the response graph of strategy profiles until converging to the stationary distribution, as is described in $\alpha$-rank. It is proved that PSRO will converge with respect to MCC, i.e. to the unique sink strongly connected components in response graph, on two-player symmetric games and, if the oracle only find strategies that are not in the population, converge on any game. Unfortunately, there is no efficient way to implement such oracle to calculate best response, so MCC is almost no better than Nash equilibrium when it comes to scalability in practical use.

\section{Counterfactual Regret Minimization}

\subsection{Regret Matching}

Counterfactual Regret Minimization (CFR) is by far the most successful family of algorithms to solve complicated extensive-form games, and also a strong weapon in the arsenal of multi-agent RL. CFR is based on regret matching algorithm, which is brought forth by Hart and Mas-Colell \cite{paper/regret_matching} to adaptively learn a correlated equilibrium by self-play. Intuitively, regret matching is an algorithm where agents analyze their history and update their policy in a "what-if" manner. For example, in a repeated rock-paper-scissor game, when an agent chooses rock and loses to paper, it will consider what would happen had it played paper or scissor, and cope with the opponent who prefers paper better, as is shown in Table \ref{table:rock-paper-scissor}. However, simply choosing best response is problematic: not only can this be exploited by a clever adversary, but it is also not practical when the calculation of best response is ineffective e.g. requires exponential time. Therefore, regret matching gives a stochastic response where the probability of actions is proportional to their cumulative positive regrets i.e. the gain of payoff had it played this action before. In the example above, the agent will play scissor twice more likely than paper in the next round, while the actions are arbitrarily chosen if there are no positive regret. It is proved that the distribution of the agents' response will converge to the correlated equilibrium when the game is played enough times.

\begin{table}
\begin{center}
\begin{tabular}{|l|l|l|l|l|}
\hline
Round & Action & Opponent & CR & ND \\
\hline
0 & N/A & N/A & (0, 0, 0) & (1/3, 1/3, 1/3)\\
\hline
1 & rock & paper & (0, 1, 2) & (0, 1/3, 2/3) \\
\hline
2 & scissor & rock & (1, 3, 2) & (1/6, 1/2, 1/3)\\
\hline
3 & rock & scissor & (1, 1, 1) & (1/3, 1/3, 1/3)\\
\hline
\end{tabular}
\end{center}
\caption{Regret matching in the rock-paper-scissor game. The cumulative regrets (CR) and next distributions (ND) are of (rock, paper, scissor).}
\label{table:rock-paper-scissor}
\end{table}

Regret matching can be viewed as an online learning algorithm where the regrets are learned to model other agents' strategies, and the bound of cumulative regret plays an important role in its performance and convergence. Hart and Mas-Colell proved in the original paper \cite{paper/regret_matching} that the cumulative regret grows sublinearly with respect to the number of rounds played. Greenwald \etal \cite{paper/BRMA} gives the bound of regret for a generalized form of regret matching, where the probability distribution of actions in the next round can be proportional to the polynomial or exponential form of cumulative regret. More theoretical results are done by Orazio \etal \cite{paper/BARMA} \cite{paper/FCRF} to use function approximators in regret matching in the context of deep learning. In the literature of reinforcement learning, however, most work is based on the original form \cite{paper/regret_matching} where the probability of actions is proportional to cumulative regret.

\subsection{CFR, CFR+ and MCCFR}

Counterfactual regret minimization (CFR) \cite{paper/regretpoker} uses regret matching algorithm to cope with extensive-form games with an exponential number of strategies with respect to steps of the game, and is born to be compatible with partially observable environment. Formally, let $R_i^{T}(I,a)$ be the regret for player $i$ in round $T$ in information set $I$ choosing action $a$, $\pi_{-i}^{\sigma^{t}}(I)$ be the probability of entering $I$ considering the strategies of every player but $i$ in round $t$, $\sigma^{t}|_{I\rightarrow a}$ be the strategy profile identical to $\sigma^{t}$ except that player $i$ always plays $a$ in $I$. Then the regret can be calculated as:
\begin{equation}
R_i^{T}(I,a)=\frac{1}{T}\sum_{t=1}^{T}\pi_{-i}^{\sigma^{t}}(I)(u_i(\sigma^{t}|_{I\rightarrow a},I)-u_i(\sigma^t,I))
\end{equation}
and the agents update their policies for the next round as:
\begin{equation}
R_i^{T,+}(I,a)=\max(R_i^{T}(I,a),0)
\end{equation}
\begin{equation}
\delta_i^{T+1}(I,a)=\left\{\begin{aligned}
&\frac{R_i^{T,+}(I,a)}{\sum_{a\in A(I)}R_i^{T,+}(I,a)} \; \text{if denominator}>0\\
&\frac{1}{|A(I)|}\qquad\qquad\qquad\text{otherwise.}\\
\end{aligned}\right.
\end{equation}
where $A(I)$ is the action set in information set $I$. It is proved \cite{paper/regretpoker} that the regret bound is $O(1/\sqrt{T})$, and is further improved to $O(1/T^{0.75})$ by Farina \etal\cite{paper/farina}.
\label{section:CFR}

In 2014, Tammelin \etal \cite{paper/CFRplus} \cite{paper/HLTH} brought forth a variant of CFR algorithm called CFR+ by substituting the regret matching algorithm in CFR by regret matching plus. The regret matching plus algorithm calculates the regret as
\begin{equation}
R_{i}^{T}(I,a)=R_i^{T-1}(I, a)+v_i(\sigma^{T}|_{I\rightarrow a},I)-v_i(\sigma^{T},I)
\end{equation}
where the counterfactual value \cite{paper/EMCFR} $v_i(\sigma, I)$ is the expectation utility of playing strategy profile $\sigma$, which is always 0 unless information set $I$ is reached.

Traditionally, due to the prohibitive time complexity for traversing the whole game tree for extensive-form games, the deployment of CFR in large-scale games such as Limited Texas Hold'em requires prior knowledge and non-trivial work for state abstraction \cite{paper/regretpoker} \cite{paper/pokeragent}. Monte-Carlo counterfactual regret minimization (MCCFR) \cite{paper/MCCFR} is proposed to reduce the time complexity, which is a sample-based variant of CFR. There are two variants of the algorithm: outcome sampling and external sampling, both of which can compute an approximate equilibrium. In outcome sampling, only a single episode of the game is sampled at each iteration, and only the nodes on the path of the game tree in this episode are updated. Since the nodes are sampled according to the opponents' strategy, there is no need to explicitly know the opponents' strategy. External sampling samples only the chance nodes (external to the player) and the opponent’s actions, which is proved to make an asymptotic improvement in computational time of equilibrium.
\begin{figure}[t]
\begin{center}
  \includegraphics[width=0.8\linewidth]{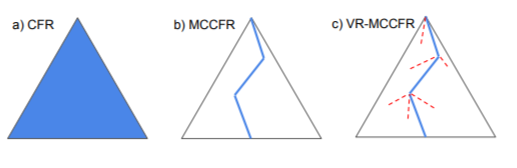}
\end{center}
   \caption{The comparison between CFR, MCCFR and VR-MCCFR. The picture is from \cite{paper/VRMCCFR}.}
\label{fig:vrmccfr}
\end{figure}

However, Monte Carlo methods are known to be of high variance. To reduce the variance of Monte Carlo method, Schmid \etal proposed a variant of MCCFR called VR-MCCFR \cite{paper/VRMCCFR} to speed up the sampling process and reduce the empirical variance. In VR-MCCFR, an action-dependent baseline is defined for each information set in a manner similar to advantage actor-critic (A2C), which allows estimates to bootstrap from other estimates within the same episode while remaining unbiased. The algorithm is also compatible with CFR+ \cite{paper/CFRplus}, and its comparison with CFR and MCCFR is shown in Figure \ref{fig:vrmccfr}. Later, Johanson \etal proposed several sampling methods for MCCFR and MCCFR+ \cite{paper/EMCFR}. The core contribution of their work is the proposal of public-chance sampling (PCS) where only the public-chance events (the uncertainty from nature) are sampled. It is proved both theoretically and empirically that PCS can significantly accelerate the training process. There are also attempts to reduce the variance by replacing Monte-Carlo by other sampling methods \cite{paper/geneCFR}. However, MCCFR is the basis of most recent works on CFR, and it is the notion of "counterfactual value" that makes Deep CFR possible.

\subsection{Deep CFR and its Applications}

As the calculation of regret becomes infeasible in large-scale extensive-form games, researchers turn to function approximators for help. The first CFR variant that uses a function approximator is the regression CFR proposed by Waugh \etal \cite{paper/RegCFR}, which uses regression tree as the function approximator. However, this algorithm has two drawbacks. First,  handcrafted features about information sets have to be manually designed for the input of the approximator, thus prior knowledge is required. Second, just as vanilla CFR, regression CFR traverses the whole game tree, which makes it unbearably costly with respect to time complexity in non-trivial games.

The milestone that marks CFR eventually joins the family of multi-agent DRL algorithms is the invention of deep counterfactual regret minimization \cite{paper/DCFR}. Deep CFR inherits both the notion of counterfactual value and the sampling method (external sampling in their method) of MCCFR. In this algorithm, a value network is trained to estimate the counterfactual value, whose goal is to approximate the regret value that tabular CFR would have produced. Meanwhile, a policy network is trained to approximate the average strategy played over all iterations, which is optional when there is enough memory to store the policy of every iteration. Later, they propose several weight functions \cite{paper/LCFR} such as linear CFR with weight $t$, and discounted CFR with weight $\frac{t^\alpha}{t^\alpha+1}$ for iteration $t$, to put more weight on the recent regret and strategy when calculating cumulative regret and the average strategy. Steinberger \cite{paper/SDCFR} simplified deep CFR and reduced the approximation error by abandoning the policy network used to approximate the average strategy.

Many variants of CFR are proposed to better integrate CFR with DRL. Peter \etal \cite{paper/RMDRL} proposed advantage-based regret minimization (ARM), which maps "information set" $I$ in game theory to "observation" $o$ in reinforcement learning, rewriting the counterfactual action-value pair as $Q_{\pi|o\rightarrow a}(o,a)$, which means the agent follows policy $\pi$ until $o$ is observed and switches to action $a$. By the approximation $v_i(\sigma|_{I\rightarrow a},I)=Q_{\pi|o\rightarrow a}(o,a)\approx Q_{\pi}(o,a)$ (notation see section \ref{section:CFR}), the clipped advantage function can be rewritten to express the clipped cumulative regret. So a Q-learning can be conducted to learn the regret value, with actions sampled in the same manner as in original CFR. Li \etal \cite{paper/DNCFR} proposed double neural CFR, which is another NN-based implementation of CFR. Two LSTM networks are trained in this algorithm, one for estimating the cumulative regret (used to derive the immediate policy) and the other for estimating an average policy.

Besides applications on traditional card games such as Limited Texas Hold'em \cite{paper/libratus} \cite{paper/poker_deepstack}, there are attempts to use deep CFR in games with deception and concealment. Serrino \etal \cite{paper/FFF} trained deep CFR agents to play Avalon, a game where players need to deduct others' identities and hide their own to achieve their goal, and achieved super-human performance with human-explainable agents, though combined with much prior knowledge.

However, as a multi-agent RL algorithm, deep CFR has two limitations. First, despite that CFR is born for partially observable environment, it requires the agent to have perfect recall i.e. recognize everything that happened in the past. Most DRL agents without recurrent neural networks (RNN) like LSTM \cite{paper/lstm}, however, do not have such ability. Second, there are games that never reaches terminal state, and agents are trained to maximize average reward instead. Such environment setting is incompatible with CFR. To solve these problems, Kash \etal \cite{paper/QCFR} proposed a Q-learning-style updating rule for agents, where agents uses the CFR algorithm locally in the transition of the Q-learning operator. The algorithm achieves iterative convergence in several games such as NoSDE, a class of Markov games proposed by Littman \etal \cite{paper/cyclic} and specifically designed to be challenging to learn, where no prior algorithm converges to an average stationary equilibrium.

\section{Conclusion}

Deep reinforcement learning has achieved outstanding results in single-agent scenarios \cite{paper/drl_atari} in recent years. However, learning in multi-agent systems is fundamentally more difficult and independently using DRL only achieves limited success in specific problems \cite{paper/rl_multiagent_DQN_pong} \cite{paper/rl_multiagent_DQN_social} \cite{paper/rl_multiagent_PPO_mojoco} \cite{paper/drl_play_ESS}. Game theory helps better understand the nature of multi-agent systems and gives much inspiration to DRL, either in theoretical concepts or algorithms, producing new techniques capable of solving multi-agent games which were once considered impossible \cite{paper/alphagozero} \cite{paper/alphazero} \cite{paper/poker_deepstack} \cite{paper/FFF} \cite{blog/openai_dota2} \cite{blog/alphastar}.

We note that none of these recent exciting results can be achieved by deep reinforcement learning or game-theoretical algorithms alone. However, previous surveys have always tried to contextualize most of multi-agent learning literature into either game theory \cite{paper/survey_mal_in_game_theory} or deep reinforcement learning \cite{paper/survey_mal_in_drl}, which makes it difficult to understand current multi-agent learning algorithms from a holistic perspective.

This survey focuses on algorithms in multi-agent systems derived from both domains, especially their evolutionary history and how they have developed from both domains. We analyze the inspiration that solution concepts give to multi-agent reinforcement learning, how fictitious self-play ends up into the toolbox of RL, and how CFR is combined with deep learning and later unified with DRL. We hope that this survey will incentivize the community of multi-agent learning to see the close connection between the domains of DRL and game theory, and motivate future research in this joint field, taking advantage of the ample and existing literature.

{\small
\bibliographystyle{ieee}
\bibliography{egpaper_final}
}

\end{document}